\documentstyle[aps,prl,epsf,floats]{revtex}

\begin{document}

\twocolumn[\hsize\textwidth\columnwidth\hsize\csname@twocolumnfalse\endcsname

\title{Discrete breathers in dc biased Josephson-junction arrays}

\author{J. J. Mazo$^{1,2}$, E. Tr\'{\i}as$^{1}$ and T. P. Orlando$^{1}$}

\address{$^{1}$ Department of Electrical Engineering and Computer Science,\\
Massachusetts Institute of Technology, Cambridge, MA 02139, USA\\}
\address{$^{2}$ Departamento de F\'{\i}sica de la Materia Condensada and ICMA \\
CSIC-Universidad de Zaragoza, E-50009 Zaragoza, Spain}

\date{\today}
\maketitle
\tightenlines

\begin{abstract}
We propose a method to excite and detect a rotor localized mode
(rotobreather) in a Josephson-junction array biased by dc currents.  
In our numerical studies of the dynamics we have
used experimentally realizable parameters and included self-inductances.
We have uncovered two families of rotobreathers.
Both types are stable under thermal fluctuations
and exist for a broad range of array parameters and sizes
including arrays as small as a single plaquette.
We suggest a single Josephson-junction plaquette as  an ideal system to
experimentally investigate these solutions.
\end{abstract}

]

\narrowtext

The phenomenon of intrinsic localization (intrinsic localized modes
or discrete breathers (DB)) is a recent discovery in the subject of nonlinear
dynamics \cite{gen}. DB are solutions to the dynamics
of discrete extended systems for which energy is exponentially localized in space.
They appear either as oscillator localized modes,
for which a localized group of oscillators librate; or rotor localized modes or rotobreathers,
for which a group of oscillators rotate while the others librate
\cite{roto}.
Recently, it has been found that DB are  not restricted to periodic
solutions but can also include more complex (chaotic) dynamics \cite{chao}.

DB have been proven to be generic solutions in hamiltonian \cite{proof}
and dissipative \cite{proof2} nonlinear lattices. It is believed 
that they might play an important role in the dynamics of a large number
of systems, such as coupled nonlinear oscillators or rotors.

Though intrinsic localized modes have been the object of great theoretical
and numerical attention in the last 10 years, they have yet to be generated
and detected in an experiment. Thus, finding the best system and method for
the generation, detection and study of an intrinsic localized mode in a
Condensed Matter system has become an important challenge \cite{JJL,Lai}.

Josephson-junctions arrays are excellent experimental systems
for studying nonlinear dynamics \cite{general}. 
In this paper we propose an experiment to detect a
rotating localized mode in JJ anisotropic ladder arrays biased by
dc external currents \cite{new}.
For this, we have done numerical simulations of the dynamics of an open
ladder including induced fields \cite{someone}
at experimentally accessible values
of the parameters of the array.
We also propose a method for exciting a
rotobreather in the array. We
distinguish between two families of solutions which present 
different voltage patterns in the array. Both types are robust to random
fluctuations and exist over a range of parameter values and array sizes.
Unexpectedly, we have found that many of the rotobreather solutions do
not satisfy
the up-down symmetry usually assumed for most types of dynamical
solutions in the ladder.  We also
show that a DB solution can be most readily studied in a single plaquette. 

According to the RCSJ model, a Josephson junction is characterized by its
critical current $I_c$,
normal state resistance $R_n$, and capacitance $C$.  The junction voltage $v$
is related to the gauge-invariant phase difference $\varphi$ as
\begin{equation}
v={\Phi_0 \over 2 \pi}{d \varphi \over dt},
\end{equation}
where $\Phi_0$ is the flux quantum.
After standard rescaling of the time by
$\tau=\sqrt{ \Phi_0 C / 2\pi I_c}$, the normalized current
through the junction is
\begin{equation}
i=\ddot{\varphi} +\Gamma \dot{\varphi} + \sin \varphi,
\label{RCSJ}
\end{equation}
here $\Gamma$ represents a damping and
is directly related to the Stewart-McCumber parameter
$\beta_c=\Gamma^{-2}=2\pi I_c C R_n^2 / \Phi_0$.

\begin{figure}[t]
\epsfxsize=2.5in
\centering{\mbox{\epsfbox{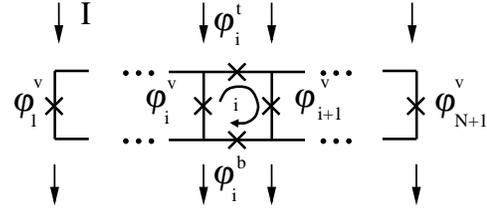}}}
\vspace{0.15in}
\caption[]{Anisotropic ladder array with uniform
current injection.  Vertical junctions
have critical current $I_{cv}$ and horizontal junctions $I_{ch}$.}
\label{fig:ladder}
\end{figure}

Our anisotropic JJ ladders (see Fig.~\ref{fig:ladder})
contain junctions of two different critical currents:
$I_{ch}$ for the horizontal junctions and $I_{cv}$ for the vertical
ones. Anisotropic arrays are easily fabricated by varying the area of the junctions.
In the case of unshunted junctions, the critical current
and capacitance are proportional to this area.
Due to the constant $I_cR_n$ product,
the normal state resistance is inversely proportional to the
junction area.
The anisotropy
parameter $h$ can then  be defined as $h=I_{ch}/I_{cv}=C_h/C_v=R_v/R_h$.

To write the governing equations of an anisotropic JJ ladder array
with $N$ cells, Fig.~\ref{fig:ladder}, we need to apply current conservation
at each node and flux quantization at each mesh.
We are including self-induced magnetic fields so that flux quantization
at mesh $j$ yields
\begin{equation}
(\nabla \times \varphi)_j=-2\pi f_j.
\label{FQ}
\end{equation}
Here $(\nabla \times \varphi)_j=
\varphi^t_j+\varphi^v_{j+1}-\varphi^b_j-\varphi^v_j$
and it represents the circulation of
gauge-invariant phase differences in mesh $j=1$ through $N$.
The self-induced flux through mesh $j$ normalized
by $\Phi_0$ is given by $f_j$.
The resulting equation can be written compactly as,
\begin{eqnarray}
& & h( \ddot{\varphi}^t_j +\Gamma \dot{\varphi}^t_j + \sin \varphi^t_j ) =
-\lambda (\nabla \times \varphi)_j \nonumber \\
& & \ddot{\varphi}^v_j +\Gamma \dot{\varphi}^v_j + \sin \varphi^v_j =
\lambda [(\nabla \times \varphi)_j - (\nabla \times \varphi)_{j-1}] + I 
\nonumber \\
& & h( \ddot{\varphi}^b_j +\Gamma \dot{\varphi}^b_j + \sin \varphi^b_j ) =
\lambda (\nabla \times \varphi)_j,
\label{eqsladder}
\end{eqnarray}
where the open boundaries can be imposed by
setting $(\nabla \times \varphi)_0=(\nabla \times \varphi)_{N+1}=0$
in Eqs.~\ref{eqsladder}. 
The system has four independent parameters: $h$, $\Gamma$,
the penetration depth $\lambda=\Phi_0/2\pi I_{cv} L$ where $L$ is the
mesh self-inductance, and the normalized external current $I$.
On writing Eqs.(\ref{FQ}) and (\ref{eqsladder}) we assume zero external field
and normalize by $I_{cv}$. Non-zero external fields can be included
in the model replacing the $(\nabla \times \varphi)_j$ terms in Eqs.(\ref{FQ})
and (\ref{eqsladder}) by $(\nabla \times \varphi)_j+ 2\pi f^{ext}_j$ where $f^{ext}_j$ is
the flux due to an applied external field, measured in terms of $\Phi_0$.

The parameter values we will consider are based on Nb-Al$_2$O$_x$-Nb
junctions  with a critical current density of $1000 \, {\rm A/cm^2}$.
Typical values of the
Stewart-McCumber parameter and the penetration depth
for arrays with $h=1/4$ are $\beta_c \sim 30$ and $\lambda \sim 0.02$.
For the purposes of this work we will let
$\Gamma=0.2$, $\lambda=0.02$ and $N=8$.

Consider the $h=0$ limit in Eqs.~\ref{eqsladder}.  In this limit the vertical junctions
behave as uncoupled damped pendula driven by an external current $I$.
There, we can think of a configuration in which one or a few of the phases are 
rotating or oscillating
around their equilibrium points while the others remain at rest. 
Thus rotor and/or oscillator localized modes appear as solutions to the dynamics when the array
is biased either by dc or ac external currents.

For a single underdamped junction driven by
a constant external current, the response measured in terms of dc
voltage presents a hysteresis loop 
between the depinning and the retrapping currents. 
In this range the
pinned ($V=0$) and rotating ($V \neq 0$) solutions coexist. Then the
rotobreather solution in the $h=0$ limit corresponds to a solution
in which the phase of one of the vertical junctions is rotating while the other
vertical junctions are at rest.

\begin{figure}[t]
\epsfxsize=3in
\centering{\mbox{\epsfbox{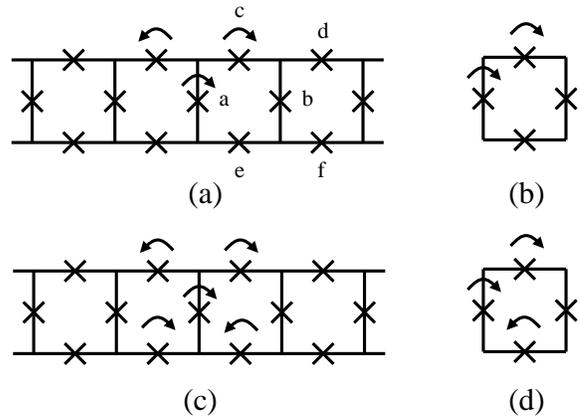}}}
\vspace{0.15in}
\caption[]{Schematic picture of rotobreather solutions:
type A rotobreather in the ladder array (a), and plaquette (b);
type B rotobreather in the ladder (c) and plaquette (d).
Arrows are associated with rotating junctions and labels in (a)
corresponds to graphs in Fig.~\ref{phaseportrait}.}
\label{redes}
\end{figure}

As $h$ is increased from zero the non-convex character of the
coupling allows for the continued existence of rotobreathers in the system.
Since a solution with a time increasing field cannot
physically exist, the flux quantization condition (Eq.~\ref{FQ}) implies that
each cell with a rotating junction must have at
least one other junction which is rotating. Thus, for the single
rotobreather solution one of the vertical and some of the horizontal 
neighboring junctions rotate.
Fig.~\ref{redes} shows schematically simple rotating localized modes in a 
ladder and in the single plaquette.
These DB
are amenable to simple experimental detection when measuring the average voltage 
through different junctions.

Although the rotobreather solution can theoretically be continued from 
its $h=0$ limit by varying $h$, 
we have developed a simple method of exciting it in
an array. This method should be experimentally reproducible and has three
steps: (i) bias all the array up to the operating point ($I=I^*$);
(ii) increase the current injected into one of the junctions to a value of the
current above the junction critical current ($I=I^* + \tilde{I}>1$);
(iii) go back to the operating point by decreasing to zero the value of this
extra current $\tilde{I}$. Typical
values of $I^*$ and $\tilde{I}$ in our simulations are 0.6. 

\begin{figure}[t]
\epsfxsize=3.5in
\centering{\mbox{\epsfbox{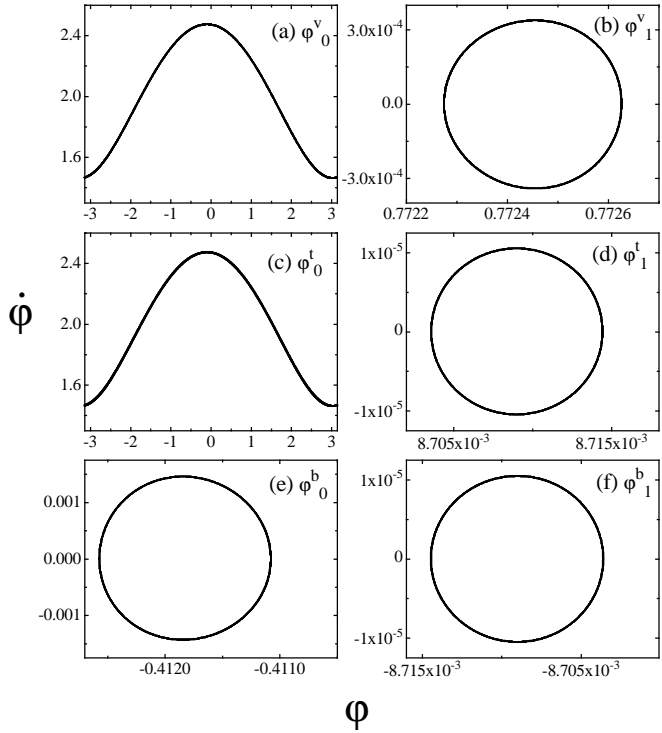}}}
\caption[]{Example of a rotobreather in an 8 cell JJ ladder
at $h=0.25$, $\lambda=0.02$; $\Gamma=0.2$ and $I_{dc}=0.6$. The only vertical
rotating junction is the central one, $j=0$, and we show the phase portrait of six of
the phases of the array: (a) $\varphi^v_0$, (b) $\varphi^v_1$, (c) $\varphi^t_0$ ,
(d) $\varphi^t_1$, (e) $\varphi^b_0$ and (f) $\varphi^b_1$ [Plotted phases are labeled
in Fig.~\ref{redes}(a)]. Also, in this case
$\varphi^v_{j}=\varphi^v_{-j}$, $\varphi^{t(b)}_{-j}=-\varphi^{t(b)}_{j-1}$.}
\label{phaseportrait}
\end{figure}

We have checked the robustness of this method under fluctuations
by simulating the equations of the ladder while adding a noise current 
to the junctions (this is the standard manner of including thermal effects in 
the system \cite{temperature}). Thus we are able to excite DB in the ladder at
some values of the parameters of the system. The solution showed in Fig.~\ref{phaseportrait} was
excited using this procedure.

Henceforth, we are going to consider ladders with an even number of cells for 
which one vertical junction (the central one) is rotating. We will relabel
this junction as $j=0$.

\begin{figure}[t]
\epsfxsize=2.8in
\centering{\mbox{\epsfbox{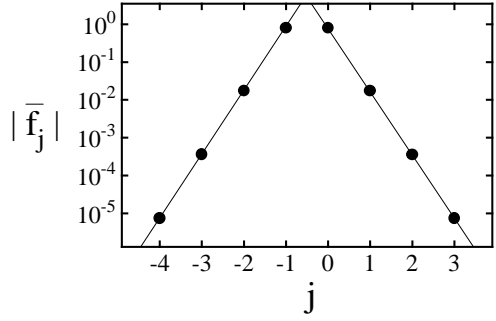}}}
\caption[]{
The average value of the induced flux ($f_{-j}=-f_{j-1}$) 
at all the cells of the ladder for 
the rotobreather shown in Fig.~\ref{phaseportrait}.
This average value decreases exponentially 
which is characteristic of DB solutions.}
\label{flux}
\end{figure}

Figs.~\ref{phaseportrait} shows a solution of a stable rotobreather in a
JJ ladder. We plot the phase portrait ($\varphi^q_j$,$\dot{\varphi}^q_j$) of some of
the superconducting gauge invariant phase differences of the array. The corresponding
junctions are shown in Fig.~\ref{redes}(a).
For clarity we have reduced the values of the phases to the $(-\pi,\pi]$
interval. We see that at this value of the penetration depth the solution is highly 
localized while three of the junctions describe a nearly sinusoidal
rotation all the others oscillate with decreasing amplitudes.
The average voltage through the three rotating junctions in the array is different from zero
and equals to zero for all the other junctions.
Fig.~\ref{flux} shows 
the average value of the induced field of
the cells of the array.
It decreases exponentially as
$\overline{f_j} \sim \mbox{e}^{-j/0.26}$
($j \geq 0$ and $f_{-j}=-f_{j-1}$)

There are some surprising characteristics in this solution.
Current conservation in the open ladder implies $i^t_j = -i^b_j$.
From Eqs.~\ref{RCSJ} and \ref{eqsladder}
we can see that $\varphi^t_j=-\varphi^b_j$ is a simple solution of
the dynamics of the array and it corresponds to the up-down symmetry of the
phases. All the previous theoretical approaches to
the dynamics of the array (which include whirling modes, resonances, row-switching, etc.)
and many of the numerical ones focus on solutions which satisfy this up-down symmetry.
However, looking at Fig.~\ref{phaseportrait} we see that the rotobreather
solution shown there does not comply with this simple symmetry; that is,
$\varphi^t_j \neq -\varphi^b_j$ although $i^t_j = -i^b_j$.  

We will distinguish between two families of single rotobreather
solutions in the ladder which present different voltage patterns.
The first family, rotobreather A [see Fig.\ref{redes}(a)], is characterized
by one vertical and two horizontal rotating junctions.  
Type A solutions have two possible configurations.
The two rotating horizontal junctions can be both in the same side,
either top or bottom, as in Fig.~\ref{redes}(a); or one  in the
top and the other in the bottom.
The second family, rotobreather B [Fig.\ref{redes}(c)], is characterized by
one vertical and four horizontal rotating junctions. The solution
shown in Fig.~\ref{phaseportrait} and \ref{flux} is a type A
rotobreather.
Up-down symmetric
solutions belong to family B but not all family B solutions satisfy this
symmetry. 

Figs.~\ref{phaseportrait} and \ref{flux} show a solution for which the scale
of localization is smaller than one cell. Thus, it is natural to study 
the DB solution in the simplest ladder array, the single plaquette.  Obviously,
the concept of exponential spatial localization is not applicable to 
the plaquette,
but all the other characteristics of the solution remain. In particular we will
also distinguish between type A and type B rotobreather solutions in the plaquette,
which in this case correspond to one vertical and one horizontal rotating
junctions [Fig.\ref{redes}(b)], and one vertical and both horizontal rotating junctions
[Fig.\ref{redes}(d)] respectively.
The single plaquette biased by dc external currents, is then proposed as the
simplest and most convenient experimental system for detecting a rotating localized mode.
The method for exciting the mode is also applicable to this system.

An important experimental issue then becomes finding the region of existence of these
DB solutions with respect to the system parameters in order to investigate the
feasibility of designing an array in to detect a DB.  To
design an array we need to calculate the junction areas, so that the
anisotropy needs to be known.
Since different values of the anisotropy affect the cell geometry, they also
change the value of $\lambda$.  On the other hand, $\Gamma$
is determined by the current density of the junctions and therefore
it is fixed and independent of the geometry while the applied 
current can be easily changed while measuring.  In order to make
an optimal design we will
fix the value of $I$ and $\Gamma$ to 0.6 and 0.2 respectively and
study DB solutions in the ($h$-$\lambda$) plane of parameters.

\begin{figure}[t]
\epsfxsize=2.5in
\centering{\mbox{\epsfbox{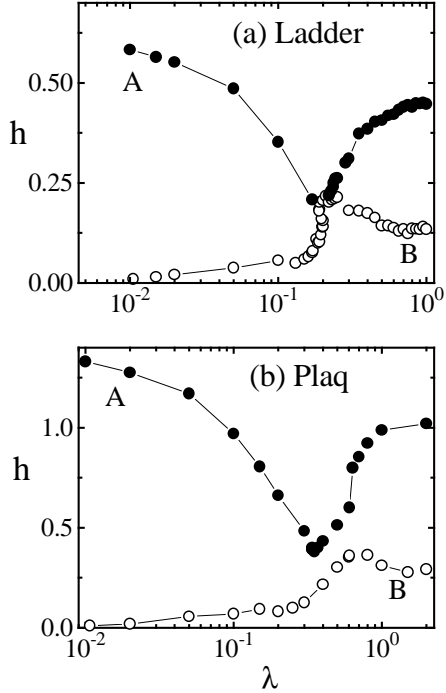}}}
\caption[]{Maximum values of anisotropy for the existence of type A
(solid circles) and type B (open circles) rotobreather solutions at
different values of $\lambda$ in an 8 cell ladder (a) and
a single plaquette array (b).
$\Gamma$ and $I$ are equal to 0.2 and 0.6 respectively.  Lines
serve as a guide to the eyes.}
\label{domains}
\end{figure}

Type A and type B rotobreathers exist close to the $h=0$ limit.
We then calculate the maximum value of the anisotropy for which a DB exists
as a stable solution of the dynamical equations for different values of $\lambda$.
Fig.~\ref{domains} (a) and (b) show the result for the ladder and the single plaquette.
The data were calculated by integrating the equations of motion
for the corresponding system with a small quantity of noise.
We start with a type B rotobreather and $h \sim 0.001$. As we increase $h$,
type B solutions
become unstable and the solution evolves to a type A rotobreather. As we further
increase $h$ this rotobreather becomes unstable and the system usually jumps to
either a pinned or a whirling state. To verify that our method is accurate, we have
calculated Floquet multipliers for periodic rotobreather solutions and found
results consistent with those shown in the Fig.~\ref{domains}.

We note that when doing this existence analysis of the solutions we find many
different single-breather solutions. Most are periodic with different periods and
amplitudes but there are some that appear to be chaotic [specially close to the
$\lambda=0.2$ region in Fig.~\ref{domains}(a)].
A detailed study of the different bifurcations which include period doubling
bifurcations to chaos is in current progress.
There also exists a large family of different multibreather solutions, each one
with its own domain of existence.

Fig.~\ref{domains} shows that at $\Gamma=0.2$ and $I=0.6$ type A solutions exist
at larger values of the
anisotropy than type B solutions. 
Also a simple inspection reveals a strong similarity
between Figs.~\ref{domains}(a) and \ref{domains}(b). The
similarity can be easily understood.
The rotobreather solution shown in Figs.~\ref{phaseportrait} and \ref{flux} presents
a mirror symmetry with respect to the rotating vertical junction:
$\varphi^v_{j}=\varphi^v_{-j}$, and $\varphi^{t(b)}_{-j}=-\varphi^{t(b)}_{j-1}$.
In the case
of solutions satisfying a mirror symmetry it is possible to
map the dynamics of a JJ ladder for which the rotating junction is the central one,
to the dynamics of a smaller JJ ladder for which the rotating junction is on one
of the ends. Then, due to the localized nature of the DB
solution 
the dynamics can be approximated by studying a single plaquette.
When doing these transformations we need to rescale two of the parameters
of the equations. Thus, results for the DB solution studied above present some
similarities with the dynamics of a DB in a single plaquette when $h_{p}=2 h_{l}$
and $\lambda_{p}=2 \lambda_{l}$.

By establishing a criteria for the design of simple experiments to detect
these intrinsic localized modes we hope to stimulate experimental 
investigations.

The research was supported in part by the
NSF Grant DMR-9610042 and DGES (PB95-0797). JJM is supported by
a Fulbright/MEC Fellowship. We thank
A.~E. Duwel, F. Falo, L.~M. Flor\'{\i}a, P.~J. Mart\'{\i}nez, and
S.~H. Strogatz for useful discussions.


\vspace{-0.2in}

\begin{references}

\vspace{-0.6in}

\bibitem{gen} A.~J. Sievers and S. Takeno, Phys. Rev. Lett. {\bf 61}, 970 (1988).
See 
Physica D {\bf 113} Issue 2-4 (1998) and Physica D {\bf 119} Issue 1-2 (1998).

\bibitem{roto} S. Takeno and M. Peyrard, Physica D {\bf 92}, 140 (1996); Phys. Rev. E
{\bf 55} 1922 (1997)

\bibitem{chao} P.~J. Mart\'{\i}nez, L.~M. Flor\'{\i}a, F. Falo and
J.~J. Mazo, to appear in Europhys. Lett.;
D. Bonart and J. B. Page, preprint.

\bibitem{proof} R.~S. MacKay and S. Aubry, Nonlinearity {\bf 7}, 1623 (1994).

\bibitem{proof2} J.-A. Sepulchre and R.~S. MacKay, Nonlinearity {\bf 10}, 679 (1997).

\bibitem{JJL} L.~M. Flor\'{\i}a, J.~L. Mar\'{\i}n, P.~J. Mart\'{\i}nez,
F. Falo and S. Aubry, Europhys. Lett. {\bf 36}, 539 (1996);
L.~M. Flor\'{\i}a, J.~L. Mar\'{\i}n, S. Aubry,  P.~J. Mart\'{\i}nez, F. Falo and
J.~J. Mazo, Physica D {\bf 113}, 387 (1998);
P.~J. Mart\'{\i}nez, L.~M. Flor\'{\i}a, J.~L. Mar\'{\i}n,
S. Aubry and J.~J. Mazo,  Physica D {\bf 119}, 175 (1998).

\bibitem{Lai} R. Lai and A.~J. Sievers, Phys. Rev. Lett. {\bf 81}, 1937 (1998).

\bibitem{general} S. Watanabe, H.~S.~J. van der Zant, S.~H. Strogatz and T.~P.
Orlando, Physica D {\bf 97}, 429 (1996); J.~C. Ciria and C. Giovannella,
J. Phys.: Condes. Matter {\bf 10} 1453 (1998).

\bibitem{new} This system has recently been proposed \cite{JJL} as a good
candidate to experimentally detect a discrete breather when biased
by ac currents. But, generating the rotobreather and the high
frequencies involved in the drive are
serious obstacles towards an experimental detection in this ac case.
Regarding the dc case, the existence of these modes in coupled rotor
systems has been proved
by R.~S. MacKay and J.-A. Sepulchre, Physica D {\bf 119}, 148 (1998)
and numerically found by
Z. Zheng, B. Hu and G. Hu, Phys. Rev. E {\bf 57}, 1139 (1998).

\bibitem{someone} The inclusion of induced magnetic fields has been shown
to be important for understanding some experimental results in
E. Tr\'{\i}as,  H.~S.~J. van der Zant, T.~P. Orlando,
Phys. Rev. B {\bf 54}, 6568 (1996); H.~R. Shea, M.~A. Itzler and M. Tinkham,
Phys. Rev. B {\bf 51}, 12690 (1995).

\bibitem{temperature} V. Ambegaokar and B.~I. Halperin, Phys. Rev. Lett.
{\bf 22}, 1364 (1969)

\end{references}
\end{document}